\documentclass[12pt,fleqn]{article}

\usepackage{latexsym}
\usepackage{amsmath}
\usepackage{amssymb}

\renewcommand{\[}{$$}

\newcommand{\rf}[1]{(\ref{#1})}

\newcommand{\ba}{\begin{array}}
\newcommand{\ea}{\end{array}}

\newcommand{\be}{\begin{equation}}
\newcommand{\ee}{\end{equation}}

\newcommand{\ods}{\par \vspace{0.5cm} \par}
\newcommand{\no}{\noindent}

\newcommand{\dis}{\displaystyle}

{\normalsize \hfill {\mbox{$\Box$}} \par  \vspace{1.5ex}}

\begin{document}

\title{\bf Modified van der Pauw method based on formulas solvable by the Banach fixed point method}
\author{Jan L.\ Cie{\'s}li{\'n}ski\thanks{e-mail:
\tt janek\,@\,alpha.uwb.edu.pl}
\\ {\footnotesize Uniwersytet w Bia{\l}ymstoku,
Wydzia\l \ Fizyki,}
\\ {\footnotesize ul.\ Lipowa 41, 15-424 Bia\l ystok, Poland}
}

\date{}

\maketitle

\begin{abstract} 
We propose a modification of the standard van der Pauw method for determining the resistivity and Hall coefficient of flat thin samples of arbitrary shape. Considering a different choice of resistance  measurements we derive a new formula which can be numerically solved (with respect to sheet resistance) by the Banach fixed point method for any values of experimental data. The convergence is especially fast in the case of almost symmetric van der Pauw configurations (e.g., clover shaped samples). 
\end{abstract}



\ods

\section{Introduction}

The van der Pauw four probe method is a standard technique for measuring the resistivity of flat thin samples of arbitrary shape \cite{Pauw1,Pauw2}. The sample have to be homogeneous, isotropic, of uniform thickness and simply connected (i.e., without isolated holes). Four contacts placed on the sample are required. They have to be geometric points located on the boundary of the sample (or, in practice, errors caused by their finite size should be sufficiently small).

The van der Pauw geometry is very popular in electric measurements and found a lot of applications in physics, compare, e.g., \cite{BO,Mo,BPEM,T...,WM}.  
The method consists in performing direct measurement of resistances $R_{12,34}$ and $R_{23,14}$ (for more details see the next section), and then using the formula
\be   \label{Pauw}
\exp \left( - \frac{ \pi d  R_{12,34}}{\rho} \right) + \exp \left( - \frac{\pi d R_{23,41}}{\rho} \right)   = 1 \ 
\ee
for computing the resistivity $\rho$ and sheet resistance $R_s = \rho/d$ \ of the sample of thickness $d$. Then, the Hall coefficient is computed as
\be
    \mu_H = \frac{\Delta R_{24,13}}{B R_s} \ ,
\ee
where $\Delta R_{24,13}$ is the change of $R_{24,13}$ due to the magnetic field $B$. Equation \rf{Pauw} is believed to be unsolvable by the fixed point method. Usually, instead of numerical procedures, a graph of the so called geometric factor is used to determine a solution of \rf{Pauw}. Some authors recommend to use tables of numerical values of this function \cite{RGA}. An inherent inaccuracy of these methods seems to be commonly recognized. 

Many attempts have been made to develop and improve the van der Pauw approach, see \cite{Ve,HTJG,Ch,WKK,KN,We,NKF,LSWH}. However, the formula \rf{Pauw} has always been treated as a starting point. 
In this Letter we will show that another formula, namely:
\be  \ba{l} \label{Pauw2} \dis
\exp \frac{ \pi d  R_{max} }{\rho}  - \exp \left( \frac{\pi d | R_{24,13} |}{\rho} \right)   = 1 \ , \\[3ex]  \dis
(\text{where} \ \ R_{max} = \max \{ R_{12,34}, R_{23,41} \}) \ , 
\ea \ee
can be used instead of \rf{Pauw}. We will show that preconditions for the Banach fixed point theorem are rigorously satisfied for any set of experimental results, usually with an excellent rate of convergence. 

Our approach is especially convenient in Hall effect measurements with symmetric (or almost symmetric) van der Pauw configuration (e.g., in the shape of a clover leaf).  Then $R_{24,13}$ is much smaller than $R_{max}$ and we need few iterations to get very accurate numerical results.

\section{A brief review of the van der Pauw method}

The main idea of the van der Pauw approach is simple and beautiful. First, one considers a sample in the form of the complex upper half plane (with contacts placed on the real axis). All computations can be explicitly done in this case. Then, one applies a deep mathematical theory (the Riemann mapping theorem) showing that any other (simply connected) sample is conformally equivalent to the upper half plane \cite{Ah}. What is more, this conformal transformation preserves all equipotential lines, current lines and boundary conditions \cite{Pauw1}. Therefore any formula which does not contain explicit information about positions of contacts is invariant with respect to such transformations, and results obtained in the case of the half plane are exactly valid for samples of arbitrary shape (provided that they have no isolated holes).   

Therefore, we consider the upper half plane, parameterized by complex coordinate $z$ (${\rm Im} z  \geqslant 0$). Four contacts are represented by $x_1, x_2, x_3, x_4$ lying on the real axis. In order to perform a measurement we inject electric current $J_{jk}$ at contact $x_j$, take it out at $x_k$ ($k\neq j$), and measure the voltage between remaining two points. Elementary considerations (based on the superposition principle) show that electric potential at $z$ is given by
\be
 \Phi (z) = \frac{ J_{jk} \rho}{ \pi d} \ln \left| \frac{z-x_k}{z-x_j} \right|  
\ee
(note that $|z_1 - z_2|$ is a distance between complex numbers $z_1$ and $z_2$). 
There are 4!=24 different ways to perform measurements described above. In any case we compute a resistance 
\be  \label{Rjkmn}
 R_{jk,m n} \equiv \frac{\Phi (x_n) - \Phi (x_m)}{J_{jk}} = \frac{ \rho}{\pi d} \ln 
\left| \frac{(x_n-x_k)(x_m-x_j)}{(x_n-x_j)(x_m-x_k)} \right|  \ ,
\ee
where $j, k, m, n$ are pairwise different (a permutation of 1, 2, 3, 4) and it is convenient to denote $R_s = \rho/d$ ({\it sheet resistance}). Thus we have 24 relations between $x_1, x_2, x_3, x_4$ and $R_s$, treated as unknowns. $R_{jk,mn}$ are calculated directly from experimental data. Eliminating $x_1, x_2, x_3, x_4$ van der Pauw obtained equation \rf{Pauw} valid for samples of arbitrary shape, compare \cite{Pauw1,Pauw2}. We stress that the exact placement of contacts on the circumference of the sample is not important with exception of their ordering. 

In the next section we study consequences of equations \rf{Rjkmn} in more detail. In particular, we derive new equation \rf{Pauw2}.

\section{Modification of the van der Pauw method}

In the formula \rf{Rjkmn} one can recognize the cross ratio, a well known and very important notion in  projective geometry. The cross ratio of four (ordered) points $x_j, x_k, x_m, x_n$ is defined as
\be \label{cross}
 (x_j, x_k; x_m, x_n) := \frac{(x_m - x_j)(x_n - x_k)}{(x_m-x_k)(x_n - x_j)} \ .
\ee
The same formula applied for a 4-tuple of complex numbers is used in conformal (M\"obius) geometry \cite{Ah,H-J}. There exists a natural generalization of the cross ratio on points in Euclidean spaces of any dimension \cite{Ci-cross}. 
\ods
Taking into account \rf{cross} we rewrite equation \rf{Rjkmn} as
\be  \label{res-cross}
\pi \dis R_{jk,mn} =   R_s  \ln | (x_j, x_k; x_m, x_n) | \ .
\ee
Cross ratios corresponding to various permutations of four points $x_1, x_2, x_3, x_4$ are related by a set of identities which can be shortly written as:
\be  \label{ident1}
 (x_j, x_k; x_m, x_n) = (x_m, x_n; x_j, x_k) = (x_j, x_k; x_n, x_m)^{-1}  , 
\ee
\be \label{ident2}
 (x_j, x_k; x_m, x_n) + (x_j, x_m; x_k, x_n) = 1 \ ,
\ee
(they can be verified by straightforward elementary calculation). In particular, on use of \rf{ident1} and \rf{ident2}  we easily derive the following equations 
\be \label{id-pauw}  (x_1, x_2; x_3, x_4)^{-1} + (x_2, x_3; x_4, x_1)^{-1} = 1 \ , 
\ee
\be \label{id-new} 
(x_1, x_2; x_3, x_4) + (x_2, x_4; x_1, x_3) = 1 \ ,  
\ee  
\be \label{id-new'} 
(x_2, x_3; x_4, x_1) + (x_2, x_4; x_1, x_3)^{-1} = 1 \ , \  
\ee 
Taking into account \rf{res-cross}, and assuming (without loss of the generality) 
\be  \label{order}
x_1 < x_2 < x_3 < x_4 \ , 
\ee
we obtain corresponding identities for resistances $R_{jk,mn}$. Equations \rf{ident1} yield the so called reciprocal and reversed polarity identities, for instance:
\be  \label{recipro}
 R_{12,34} = R_{34,12} = R_{21,43} = R_{43,21} \ .
\ee
They are useful for eliminating some side effects (one takes an average of the above four measurements insted of $R_{12,34}$, etc.). In our approach improvements of this kind  can be done in  exactly the same way as in the standard van der Pauw method. 
Note that ordering \rf{order} means that contacts $x_1, x_2, x_3, x_4$ are placed in exactly this order (counterclockwise) on the circumference of the sample.

Cross ratios are not necessarily positive. Using \rf{cross} and \rf{order} we can determine signs of cross ratios. Moreover, equation \rf{id-pauw} implies upper bounds on  both (positive) components. Thus: 
\be  \ba{l} 
 (x_1, x_2; x_3, x_4) > 1 \ , \\[1ex]
(x_2, x_3; x_4, x_1) > 1 \ , \\[1ex]
(x_2, x_4; x_1, x_3) < 0 \ .
\ea \ee
Equation \rf{id-pauw} yields van der Pauw's formula \rf{Pauw}. 
Surprisingly enough, equations \rf{id-new}, \rf{id-new'} lead to  new, physically meaningful, formulas:
\be  \label{moje}
  \exp ( \pi R_{12,34}/ R_s)  - \exp ( \pi  R_{24,13}/ R_s)  = 1 .   
\ee
\be  \label{moje2}
  \exp ( \pi R_{23,41}/ R_s)  - \exp ( - \pi  R_{24,13}/ R_s)  = 1 .   
\ee
For further analysis we choose the first equation if $R_{24,13} > 0$ or the second equation if $R_{24,13} < 0$. In the first case we have $R_{12,34} > R_{23,41} > 0$, while in the second case $R_{23,41} > R_{12,34} > 0$. Both cases can be shortly represented as equation \rf{Pauw2} where $R_{max}$ denotes greater of two values: $R_{12,34}$ or $R_{23,41}$.

\section{Fast converging numerical iterations}

Equation \rf{Pauw2} can be rewritten as: 
\be \label{modPauw}
 x = \ln \left( 1 + e^{k x} \right) \ , \quad  k = \frac{|R_{24,13}|}{R_{max} } \ ,
\ee
where $x = \pi R_{max}/ R_s$. The discussion at the end of the previous section shows that $0 \leqslant k < 1$. 

  Equation \rf{modPauw} has a form $x = F (x)$, characteristic for the Banach fixed point method. In order to obtain a  solution (the fixed point of the map $F$) one has to iterate: $x_{n+1} = F (x_n)$.  
We are going to show that function $F (x) = \ln \left( 1 + e^{k x} \right)$ satisfies preconditions for the Banach fixed point theorem (for any  $k$). Indeed, $F$ maps segment $L_k = \left[ \ln 2, \frac{\ln 2}{1-k} \right]$ into itself because: 
\be  \ba{l}  \dis
 x \geqslant \ln 2 \ \Rightarrow \ F (x) \geqslant \ln ( 1 + 2^k ) \geqslant \ln 2 \ , \\[2ex] \dis
x \leqslant \frac{\ln 2}{1-k} \ \Rightarrow \ F (x) \leqslant \frac{k \ln 2}{1-k} + \ln 2 = \frac{\ln 2}{1-k} \ ,
\ea \ee 
where we took into account \ $F (x) = k x + \ln (1 + e^{- k x})$. Then, 
\be
  |F'(x)| = \frac{k}{1 + e^{- k x}} \leqslant k \  
\ee
for any $x \in L_k$. 
Therefore, by virtue of Lagrange's mean value theorem 
\be
 \frac{| F (x_1) - F(x_2)|}{|x_1- x_2|} = |F' (c)| \leqslant k < 1 
\ee
(for any $x_1, x_2 \in L_k$) which means that $F$ is a contraction of the segment $L_k$.  

In order to estimate the number of iterations $N$ needed to obtain a prescribed accuracy $\delta$ we require that the length of the segment after applying $N$ contractions is smaller than $\delta$: 
\be  \label{est}
   \frac{k^{N+1} \ln 2}{1-k} \leqslant \delta \quad \Rightarrow \quad N \approx  
\frac{\ln \left( \frac{(1-k) \delta}{k \ln 2} \right)}{\ln k } \ .
\ee 
The actual number of iterations is, of course, much smaller.  Table~\ref{Tab-k} shows the number of iterations needed to obtain the accuracy $\delta = 10^{-5}$. For $k$ approaching 1 the number of iterations increases (tending to infinity). In this region ($k \approx 1$) it is better to use another iterating scheme, see below.
Note that as an initial point we took $x_0 = \ln 2$ (this is almost obligatory for small $k$, when the length of segment $L_k$ is very small and only $x_0=\ln 2$ belongs to any $L_k$).  Table~\ref{Tab-k} contains also corresponding values of the relative sheet resistance ${\hat R}_s$ defined by 
\be
   {\hat R}_s = \frac{R_s}{R_{max}} = \frac{\pi}{x} \ ,
\ee
where $x$ is the solution of \rf{modPauw}. 

Multiplying equation $e^x= 1 + e^{kx}$ (equivalent to \rf{modPauw}) by $e^{-x}$ we get: $e^{-x} = 1-e^{kx - x}$. Hence we have another form of equation \rf{modPauw}: 
\be  \label{modPauw2} 
  x = - \ln (1 - e^{-k'x}) \ , \quad k'=1-k \ . 
\ee
One can rigorously show that preconditions for the Banach fixed point method are satisfied (at least for sufficiently small $k'$, namely $k' < 0.125$) provided that as a starting point we take $x_0 = - \ln k'$ (in practice, the Banach method seems to work very well for larger range of $k'$, at least up to $k' \approx 0.25$). We omit technical details. Instead, we present Table~\ref{Tab-k'} showing that for small $k'$ (i.e., $k \approx 1$) equation \rf{modPauw2} is excellently solvable by the fixed point method.

\renewcommand{\arraystretch}{1.3} 
 
\begin{center}
\begin{table} \caption{ Number of iterations $N$ necessary to obtain solution $x$ of Eq.\ \rf{modPauw} ($x_0 = \ln 2$, $\delta = 10^{-5}$) and $ {\hat R}_s = R_s/R_{max}$ as a function of $k = |R_{24,13}|/R_{max}$.  } \label{Tab-k}
\par \vspace{0.3em} \par 
\small
\begin{tabular}{|c|c|c|c|c|c|c|c|c|c|c|c|c|c|}  \hline
$k$ & 0 & 0.1 & 0.2 & 0.3 & 0.4 & 0.5  & 0.6 & 0.7 & 0.8 &  0.9  \\ \hline   
$N$ & 1 & 4 & 5 & 6 & 7 & 9 &  12 & 15 & 24 & 42  \\ \hline 
${\hat R}_s $ &  4.532   &  4.302   &  4.062  &  3.811  &  3.546   &  3.264   & 2.960   &  2.623 &  2.234  &  1.743   \\ \hline
\end{tabular}
\end{table}

\begin{table} \caption{ Number of iterations $N$ necessary to obtain solution $x$ of Eq.\ \rf{modPauw2} ($x_0 = - \ln k '$, $\delta = 10^{-5}$) and ${\hat R}_s = R_s/R_{max}$ as a function of $k '=1-k$.  } \label{Tab-k'}
\par \vspace{0.3em} \par 
\small
\begin{tabular}{|c|c|c|c|c|c|c|c|c|c|c|c|c|c|}  \hline
$k'$ & 0.2 & 0.1  & 0.01 & $10^{-3}$ & $10^{-4}$ &  $10^{-6}$ & $10^{-8}$ & $10^{-10}$ & $10^{-12}$ &  $10^{-15}$ \\ \hline   
$N$ & 22 & 18  & 10 & 8 & 7 & 5 &  4 & 4 & 4 & 3  \\ \hline 
${\hat R}_s $ &  2.234   &  1.743   &  0.924  &  0.598  &  0.434   &  0.276  & 0.200   &  0.157  &  0.129  &  0.101   \\ \hline
\end{tabular}
\end{table}
\end{center}

{\small 

\section{Summary}
In this Letter we proposed an alternative approach to the standard van der Pauw method. Measurements are essentially the same as in the standard method and produce three resistances: $R_{12,34}$, $R_{23,41}$, $R_{24,13}$ (reciprocal and reversed resistances can be used for improving the accuracy, compare \rf{recipro}). We take $R_{24,13}$ and greater of  remaining two resistances, denoting it by $R_{max}$.  Then we compute two coefficients: $k = |R_{24,13}|/R_{max}$ and $k'=1-k$. In order to find the sheet resistance we solve either \rf{modPauw} (for $0 \leqslant k  < 0.9$) or \rf{modPauw2} (for $0.8 < k < 1$) and calculate $R_s = \pi R_{max}/x$. In the indicated ranges of $k$ both equations are solvable by the Banach fixed point method with excellent rates of convergence. 
 \ods

\no {\bf Acknowledgments.} I am grateful to Kamil {\L}api{\'n}ski for turning my attention on the van der Pauw method and to Krzysztof Szyma{\'n}ski for discussions.   
}

\end{document}